# Optical Semantic Communication through Multimode Fiber: From Symbol Transmission to Sentiment Analysis

ZHENG GAO, TING JIANG, MINGMING ZHANG, HAO WU,* AND MING TANG*

## Affiliations

Wuhan National Laboratory for Optoelectronics, Next Generation Internet Access National Engineering Laboratory, and Hubei Optics Valley Laboratory, School of Optical and Electronic Information, Huazhong University of Science and Technology, Wuhan 430074, China

## Email addresses

zhenggao@hust.edu.cn; jtjt@hust.edu.cn; carlzhang@hust.edu.cn; wuhaoboom@hust.edu.cn; tangming@mail.hust.edu.cn;

## Contact details

*Corresponding authors: Hao Wu (0086-15827405305); Ming Tang (0086-18607128843)

## Author contributions

M. T. proposed the idea and the conceptual design. Z. G. and T. J. built the experiments, optimized the system, and performed the measurements. M. M. Z. helped the experiments. Z. G., T. J. and H. W. wrote the manuscript. All authors revised the manuscript.

**Abstract**

We propose and validate a novel optical semantic transmission scheme using multimode fiber (MMF). By leveraging the frequency sensitivity of intermodal dispersion in MMFs, we achieve high-dimensional semantic encoding and decoding in the frequency domain. Our system maps symbols to 128 distinct frequencies spaced at 600 kHz intervals, demonstrating a seven-fold increase in capacity compared to conventional communication encoding. We further enhance spectral efficiency by implementing 4-level pulse amplitude modulation (PAM-4), achieving 9.12 bits/s/Hz without decoding errors. Additionally, we explore the application of this system for sentiment analysis using the IMDb movie review dataset. By encoding semantically similar symbols to adjacent frequencies, the system's noise tolerance is effectively improved, facilitating accurate sentiment analysis. This work highlights the potential of MMF-based semantic communication to enhance both capacity and robustness in optical communication systems, offering promising applications in bandwidth-constrained and noisy environments.

## 1. Introduction

The increasing demand for communication capacity presents significant challenges for conventional communication systems. Semantic communication, an emerging technology, has garnered considerable attention for its potential to enhance transmission efficiency and robustness [1-3]. Unlike conventional communication systems that prioritize data fidelity and the accurate reception of individual symbols or bits, semantic communication emphasizes the precise transmission of semantic information. This approach facilitates purposeful information exchange and optimizes communication resources for efficient transmission.

The transmission of semantic information is vital for semantic communication systems. Fading and noise in the physical transmission channel of optical communication can cause the loss of transmitted semantic information, leading to semantic distortion. The current research trend is co-designing source compression and channel coding to mitigate transmission errors. To enhance robustness against optical link impairments, convolutional neural network (CNN) is introduced into the semantic decoding network to implemented joint optimization [4]. Another deep learning-enabled system employs joint semantic-channel coding to address channel noise and semantic distortion [5]. However, these solutions are primarily focused on optimizing channel encoding and decoding, with minimal connection to the semantic information behind the bits. In fact, the transmission system at the physical layer can also be adapted according to the requirements of semantic communication.

To address the limitations of current approaches, it is essential to explore new physical layer solutions that can better support the unique requirements of semantic communication. One promising direction is to map semantic symbols to optical frequencies for transmission through fibers. To achieve dense frequency multiplexing and identification, we leverage the unique properties of multimode fibers (MMFs). Recently, multimode fibers (MMFs) have demonstrated high precision and cost-effective advantages in spectral analysis. The sensitivity of MMF modes to optical frequency enables high-precision wavelength identification, with spectral resolution reaching the femtometer even attometer levels [6-8].

However, most MMF spectrometers use cameras to capture speckle patterns, which limits the detection speed in high-speed communication systems. To address this limitation, we previously used multicore fiber (MCF) to replace the camera and achieve ultrafast wavelength detection, with a detection speed of 100 MHz [9]. Given their high-speed spectral analysis capabilities, we propose that MMFs can also be employed in optical communication systems for both transmission and spectral analysis.

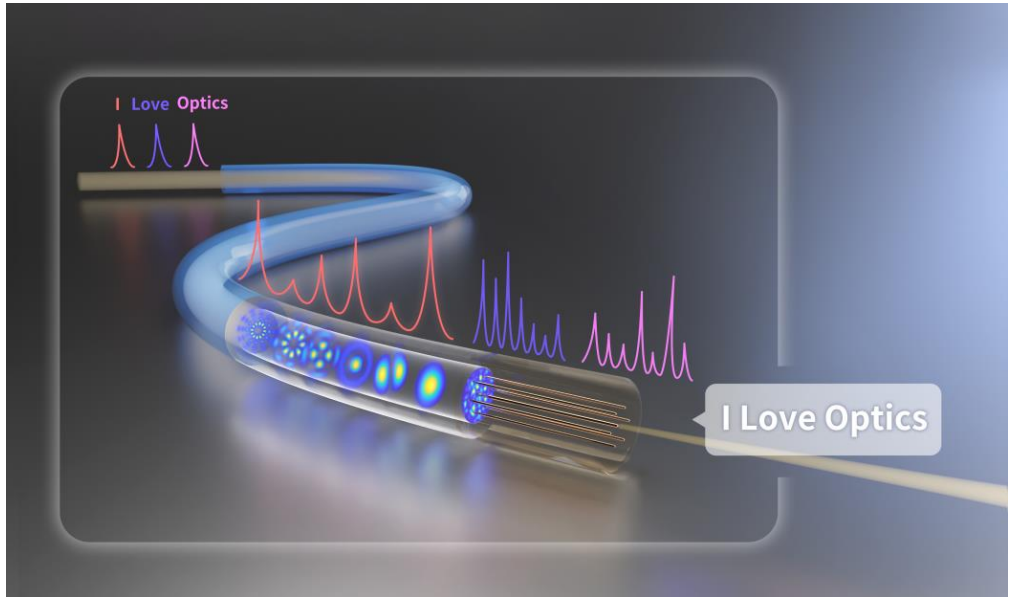

**Fig. 1. Propagation of light pulses with intermodal dispersion in MMF**. Words are mapped to different frequencies (different colors), which are then modulated onto optical pulses. During propagation in the multimode fiber (MMF), the optical pulses are stretched into different temporal dispersion curves. This allows for the identification of the pulse frequencies at the receiver, thereby enabling the reconstruction of the input words.

In this paper, we propose an optical semantic communication transmission system based on MMFs, as schematically shown in Fig. 1. Symbols or semantic messages are mapped to different frequencies, which are then modulated onto optical pulses. The intermodal dispersion in MMFs broadens optical pulses and stretches the input signal into a sequence of sub-pulses, each corresponding to a different transmission mode. By utilizing dense and independent optical frequency as the carrier for signal encoding, we achieve efficient message transmission. At the receiver end, the speckle patterns generated by the MMF are sampled at high-speed using multicore fiber (MCF). This sampling process converts the spatial distribution of the modes into single-pixel temporal dispersion curves. Each temporal dispersion curve acts as a fingerprint, representing a specific incident light frequency and its corresponding semantic symbol, thereby enabling the reconstruction of the input messages.

We experimentally demonstrate that by mapping symbols to 128 distinct frequencies spaced at 600 kHz intervals, the symbol transmission based on multimode fibers increases the capacity by seven-fold compared to conventional communication encoding. Additionally, by varying the voltage of the intensity modulator in our experiments, we show that 4-level pulse amplitude modulation (PAM-4) does not result in any decoding errors, increasing the spectral efficiency by two times to 9.12 bits/s/Hz. Furthermore, we explore the application of MMF-based semantic communication for sentiment analysis using the IMDb movie review dataset. Our approach employs an LSTM-based deep learning model to analyze sentiments, and we introduce a novel semantic encoding method that enhances transmission robustness. This method demonstrates robust performance even under high symbol error rate conditions.

## 2. Principles

*2.1 Intermodal Dispersion & Spatial speckle*

In MMFs, multiple orthogonal modes are transmitted concurrently. Variations in group velocities among these modes lead to intermodal dispersion, stretching the input optical pulse into a series of independent sub-pulses in the time domain, each corresponding to a different

transmission mode. The intensity of the optical field at the output plane of the MMF can be expressed as

$$E_{Out} = \sum_m A_m e^{i(\beta_m z - \omega(t - \Delta\tau_m))} F_m(x, y) \qquad (1)$$

where $E_{Out}$ is the electric field intensity, $A_m$ is the amplitude of the m-th mode, $\beta_m$ is the propagation constant, $\omega$ is the angular frequency of light, $\Delta\tau_m$ is the group delay, and $F_m(x, y)$ is the spatial field distribution function. The maximum pulse broadening due to intermodal dispersion can be expressed as [10]

$$\Delta\tau = \frac{L}{c} \cdot \frac{\Delta n}{n} \qquad (2)$$

where $L$ is the length of the fiber, n is the average refractive index of the fiber and $\Delta n$ is the difference in effective refractive index between the fastest and slowest modes in the fiber. These equations assume each mode transmits independently without interference. However, practical MMFs may experience mode coupling due to imperfections like microbends, stress, and manufacturing defects, affecting mode amplitude and phase [11]. When the MMF remains fixed, slowly varying environmental factors can be considered negligible over short periods, allowing the transmission matrix to be viewed as deterministic and invariant [12]. Under this assumption, the intensity distribution of the dispersed sub-pulse sequence will indicate the impact of changes in the incident optical frequency on the propagation constant. This characteristic provides a mapping relationship between the spectrum and the sub-pulse sequence, enabling reconstruction of the incident optical frequency through time-domain detection of the single-pixel pulse sequence.

In addition to differences in group velocities, each mode in an MMF has a unique spatial distribution. The speckle pattern of the MMF is the superposition of the intensities of these spatial modes. Typically, cameras are used to capture the intensity information of the speckle pattern, which is then used to analyze the incident optical field or calculate the intensity of each mode. However, the frame rate limitation of the camera usually prevents capturing time-domain differences in the modes' group velocities. The spatial distribution characteristics of the modes have the potential for multiplexing in the spatial dimension. In our approach, not all spatial location information is necessary for spectral reconstruction, meaning the speckle pattern contains a large amount of redundant information. This redundancy can be utilized for message distribution or to enhance the robustness of signal recovery.

### 2.2 Encoding and decoding methods

In our system, the encoding process involves a predefined mapping between symbols and optical frequencies, as schematically shown in Fig. 2. A specific set of optical frequencies is selected, each corresponding to a particular encoding symbol (e.g., letters, words, or semantic information). This selection and the encoding relationship can be adjusted based on communication requirements. The number of available channels for optical frequency encoding is adjustable, constrained by the available bandwidth, frequency spacing, and data processing algorithms.

The dispersion curves at the receiver act as fingerprints of the incident light frequency. During initialization, a sequence of optical frequencies containing all possible encoded information is generated as a fingerprint dictionary. Frequency recognition is performed by matching each pulse to the closest fingerprint in the dictionary using a correlation coefficient algorithm. The symbols are then decoded based on the encoding relationship between symbols and optical frequencies.

### 2.3 Single-Sideband Frequency Modulation

In our experiments, single-sideband modulation (SSB) is employed to rapidly shift the incident frequency with an in-phase/quadrature (I/Q) modulator, which consists of two Mach–Zehnder intensity modulators with a π/2 phase shifter between the two arms, biased at the null point. he optical field at the modulator output is given by

$$E_{out} = \frac{1}{2} E_{in} \cdot [sin(V_1 \cdot \frac{\pi}{V_\pi}) + j\, sin(V_2 \cdot \frac{\pi}{V_\pi})] \quad (3)$$

where $E_{in}$ is the input optical field, $V_1$ and $V_2$ are the two voltages used to drive the I and Q arms, and $V_\pi$ is the voltage required for the transfer function to change from the minimum to the maximum for each intensity modulator. We loaded the cosine and sine signals of frequency $f_m$ onto the two arms of the I/Q modulator. When operating in the linear range, the output optical field can be approximately expressed as

$$E_{out} \approx E_{in} \frac{\pi(V_1 + jV_2)}{2V_\pi} = E_0\, exp(j2\pi(f_0 + f_m)t) \cdot \pi V_0 / 2V_\pi \quad (4)$$

where $E_0$ is the amplitude of the input optical field, $V_0$ is the voltage amplitude of the signal, $f_0$ is the optical frequency of the input light, and $f_0 + f_m$ is the required frequency. The I/Q modulator is controlled by an arbitrary waveform generator which generates high-speed light pulses with controlled wavelength variation signals.

## 3. Experimental setup

The experimental setup for the proposed MMF-based frequency-encoded optical communication system is illustrated in Fig. 2. A fiber laser with a 1 kHz linewidth generates continuous monochromatic light. An arbitrary waveform generator (AWG) and an I/Q modulator are used to adjust the frequency of the incident light. The intensity modulator then converts the continuous light into a pulsed sequence. An erbium-doped optical fiber amplifier (EDFA) compensates for the loss from the modulators. The single-mode fiber is then spliced into a step-index MMF (1km length, 105/125 µm core/cladding diameter), with the other end fused to a seven-core MCF. The MCF comprises 7 single-mode fiber cores with a cladding diameter of 150 µm and core pitch of 42 µm. Light pulses from one core of the MCF are detected by a single photodetector and collected by a digital storage oscilloscope.

In our experiment, chromatic dispersion can be ignored over the 1km transmission distance. The optical power is well below the threshold that would induce nonlinear effects. To avoid overlap between pulses broadened by intermodal dispersion, we set the signal pulse rate to 50 MHz and the pulse width to 0.05 ns.

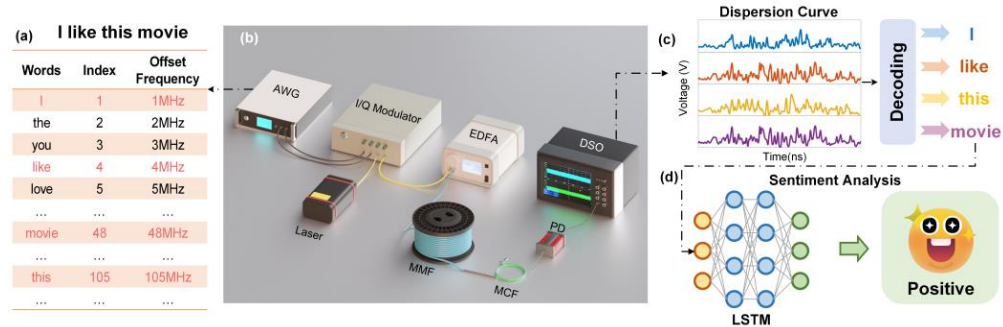

**Fig. 2.** (a) Encoding: Each word is mapped to an optical frequency according to the encoding table. (b) Schematic of the experimental setup. AWG: Arbitrary waveform generator. (c) Decoding: Recognize each dispersion curve as the corresponding frequency using the fingerprint dictionary, and then reconstruct the words based on the encoding table. (d) Implement sentiment analysis using a deep learning model.

## 4. Results and Discussion

### 4.1 ASCII encoding transmission

To validate our system's performance in practical transmission, we selected 128 independent frequencies with identical intervals, corresponding to the 128 symbols (7 bit) in the ASCII table. A one-hot encoding scheme was used between symbols and optical frequencies. The optical pulses were modulated to the corresponding frequencies based on the text content to be transmitted. At the receiving end, the text information was decoded using the fingerprint information of the dispersion curves of each pulse, as detailed in Principles section. This method can be considered as a 128-channel frequency-shift keying (FSK) system with a narrow occupied spectral width.

Figure 3 illustrates the dispersion curves for input optical pulses at various frequencies. The variation of dispersion curves is positively correlated with the frequency. Hence, the closer the frequencies of two incident pulses, the more similar their dispersion curves at the receiver, making them harder to distinguish. The accuracy of symbol reconstruction is closely related to the frequency interval during encoding. While reducing the frequency interval can further compress the occupied spectral bandwidth, it also increases the symbol error rate. Figure 4 presents the relationship between the communication symbol error rate and the frequency interval. We experimentally achieved zero-error transmission with a frequency interval of 600 kHz, equivalent to a wavelength interval of 5 fm. The symbol rate of our system is 50 MHz, leading to a spectral efficiency of 4.56 bits/s/Hz, as defined below.

$$\eta = \frac{R_b}{B} = \frac{50 M/s \times 7 bit}{600 kHz \times 128} = 4.56 \, bit/s/Hz \tag{5}$$

The intensities of output dispersion curves vary linearly with the intensity of the incident optical pulses. We tested the feasibility of further increasing capacity by implementing amplitude modulation in addition to frequency modulation. By varying the voltage of the intensity modulator in our experiments, we demonstrated that PAM-4 did not result in any decoding errors, increasing the spectral efficiency by two times to 9.12 bits/s/Hz. The intensity difference of the PAM-4 dispersion curve is clearly distinguishable, as shown in Fig. 5.

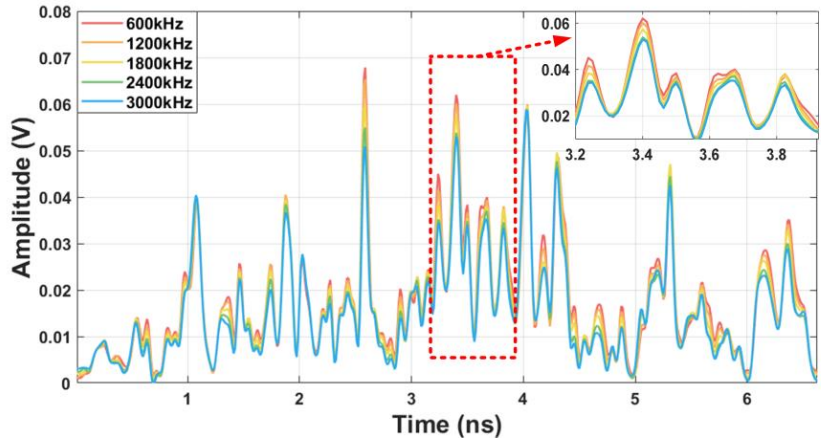

**Fig. 3.** Dispersion curves for optical frequencies offset by 600 to 3000 kHz relative to the central wavelength of 1550 nm. The variation of dispersion curves is positively correlated with the frequency.

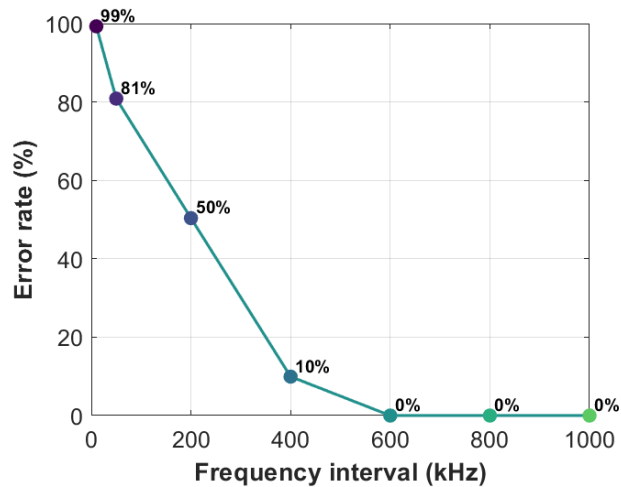

**Fig. 4.** Transmission symbol error rate of different frequency intervals.

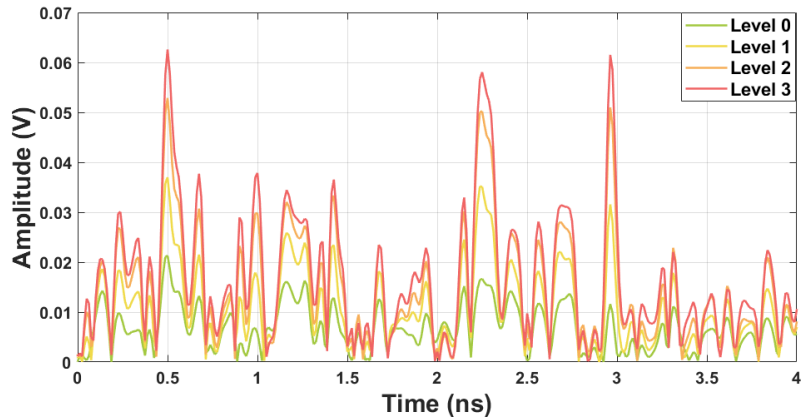

**Fig. 5**. Dispersion curves for PAM-4 experiment. The incident pulses at the same optical frequency are modulated with 4-level intensity, exhibiting distinct differences between the dispersion curves.

*4.2 Words encoding transmission and sentiment analysis*

4.2.1 IMDb and LSTM

Semantic communication is an emerging paradigm that enhances communication efficiency and noise resistance by transmitting semantic information rather than traditional bit information. In this section, we explore the application of MMF-based semantic communication in sentiment analysis using the movie review dataset provided by the Internet Movie Database (IMDb).

IMDb is a renowned online database for movies, TV shows, and celebrity information, offering extensive data including user ratings and reviews. Due to its vast amount of text data and clear sentiment labels, the IMDb movie review dataset plays a significant role in sentiment analysis, particularly within the fields of natural language processing (NLP) and machine learning. The dataset typically includes review texts and corresponding sentiment labels indicating whether the reviews are positive or negative. Consequently, IMDb has become a standard benchmark for sentiment analysis. The specific process is shown in Fig. 6.

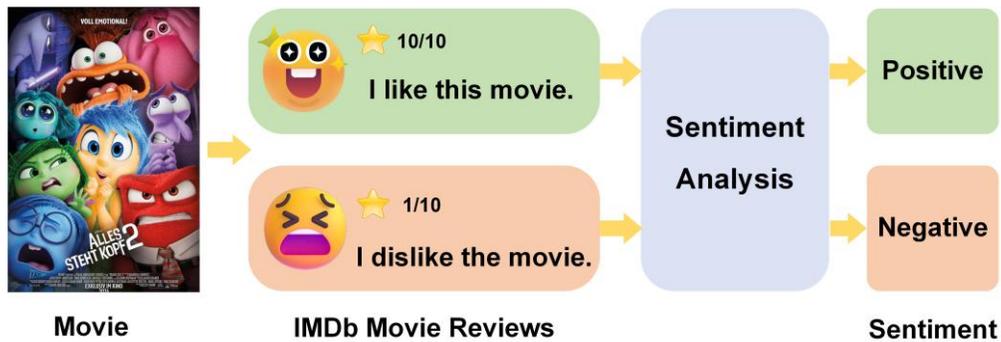

Fig. 6. Sentiment analysis diagram based on IMDb. The overall process is based on user reviews of movies from the IMDb website to achieve the analysis of positive and negative sentiments.

We utilized the IMDb movie review dataset provided by the Keras deep learning library. This dataset consists of 50,000 movie reviews from the IMDb website, split evenly into 25,000 reviews for training and 25,000 reviews for testing. The table below presents 10 selected movie review sentences from the IMDB dataset, annotated with their respective sentiments (positive or negative). Each sentence is assigned a sentiment label, where "0" indicates a negative sentiment and "1" indicates a positive sentiment. These sentences are used to provide a clear binary classification task for sentiment analysis.

Table. 1. Example movie reviews and sentiment labels from the IMDB Dataset.

| Movie Reviews | Sentiment | Label |
| --- | --- | --- |
| A rating of 1 does not begin to express how dull, depressing, and relentlessly bad this movie is. | Negative | 0 |
| No comment. Stupid movie. Acting average or worse. Screenplay makes no sense at all. Skip it. | Negative | 0 |
| Long, boring, blasphemous. Never have I been so glad to see ending credits roll. | Negative | 0 |
| I wouldn't rent this one even on dollar rental night. | Negative | 0 |
| I hope this group of filmmakers never reunites. | Negative | 0 |
| Absolutely fantastic. Whatever I say wouldn't do this underrated movie the justice it deserves. Watch it now. fantastic! | Positive | 1 |
| A touching movie. It is full of emotions and wonderful acting. I could have sat through it a second time. | Positive | 1 |
| Add this little gem to your list of holiday regulars. It is sweet, funny, and endearing. | Positive | 1 |
| I don't know why I like this movie so much, but I never get tired of watching it. | Positive | 1 |
| Brilliant and moving performances by Tom and Peter Finch. | Positive | 1 |

The overall sentiment analysis process is shown in Fig. 7. During the data reading phase, we limit the vocabulary size to the most frequent words of 10000, which helps to manage the complexity of the model while maintaining a high level of accuracy. Additionally, we standardized the length of all reviews to 500 words by truncating longer reviews and padding shorter ones. Through this preprocessing step, we obtained both the training and test sets, along with corresponding labels indicating the sentiment polarity (positive or negative) of each review. Fig. 7. (a). represents the data preprocessing section, which involves tokenization and indexing. The tokenization process involves converting each review into a sequence of tokens, where each token represents a word of text. After tokenization, the next step is indexing, where the sequences of tokens are converted into a sequence of integers using the word index. By following these steps, we ensure that the text data is properly preprocessed and standardized, facilitating effective training and evaluation of the sentiment analysis model.

To conduct sentiment analysis on the IMDb movie review dataset, we designed and implemented a deep learning model based on Long Short-Term Memory (LSTM) networks. The detailed network structure is shown in Fig. 7. (b). The model architecture includes an embedding layer, an LSTM layer, and an output layer. The embedding layer maps input word indices to a 64-dimensional vector space, converting words into continuous vector representations. The LSTM layer, with 64 units, employs a 10% recurrent dropout and L2 regularization (coefficient of 0.001) to reduce overfitting. The output layer is a fully connected layer with a single neuron and a sigmoid activation function, used to output the sentiment polarity of the reviews. The model was compiled using the RMSprop optimizer and binary cross-entropy as the loss function, with accuracy as the performance metric. During training, we set the batch size to 256. After training, the model was evaluated on an independent test set to determine its generalization capability.

Word embedding is a technique that involves mapping words from a high-dimensional space (where each word is represented as a unique index or one-hot vector) to a lower-dimensional continuous vector space. We implement this transformation process using the embedding layer of a neural network, as shown in Fig. 7. (c). The embedding layer is a trainable layer that converts input words (represented as integer indices) into dense embedding vectors of fixed dimensions. The embedding layer learns the embedding parameters during the training process, adjusting the parameters to minimize the prediction error. Through training, word embeddings can capture semantic similarities and grammatical relationships between words.

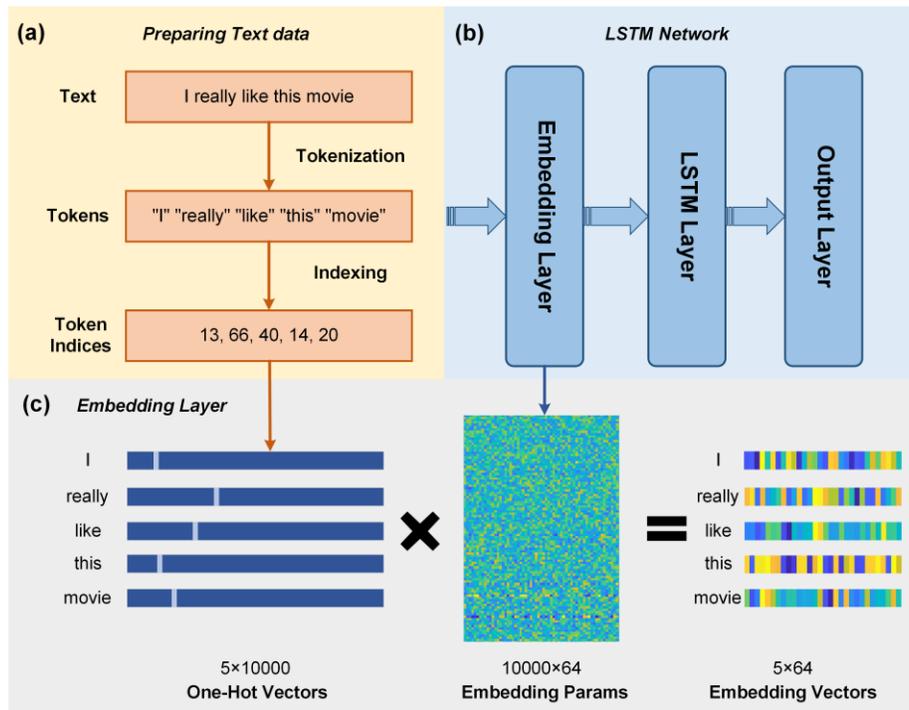

Fig. 7. The overall sentiment analysis process. (a) Text data preparation: Converting text into sequences of tokens and indexing them for model input. (b) The detailed LSTM network structure. (c) Embedding layer principle in one sentence: an embedding layer transforms high-dimensional sparse word indices into dense, low-dimensional vectors that capture semantic relationships between words.

4.2.2 Semantic Encoding Rules

In this section, we present a novel semantic encoding method designed to improve the robustness and accuracy of semantic transmission in MMF-based systems. Fig. 8 shows the histogram of symbol estimation errors at a error rate of 43% in word transmission system. Even with a high overall symbol error rate, most estimation errors deviate by only one symbol (from target symbols to adjacent symbols), with the maximum deviation being within a range of 4 symbols. This is because we use MMF-based FSK encoding for transmission. When the difference in symbol indices is small (narrow wavelength spacing), the resulting modal dispersion time sequences are more similar and harder to distinguish. Therefore, when there is an error in decoding the symbol index, the erroneous symbol index tends to become the adjacent or nearby symbol index.

In the original IMDb word indexing vocabulary encoding method, indices are based on word frequency, with no inherent semantic relationship between adjacent indices. If we transmit according to this word frequency encoding rule, it does not enhance semantic robustness. Based on the error characteristics of our MMF semantic transmission system and the semantic properties of the embedding layer, we established a new encoding rule to index word tokens. Using the pre-trained embedding layer's weight parameters, we obtained the word embedding vector representations for the top 10,000 words. We used cosine similarity to measure the similarity of word embedding vectors, where a higher cosine similarity indicates greater semantic similarity between two words. We proposed a new encoding method that encodes indices based on the semantic similarity of word embeddings, using a greedy algorithm to maximize the cosine similarity of word vectors between adjacent indices. Through this method, the encoded indices better reflect the semantic relationships between words, thereby improving semantic robustness and accuracy during transmission.

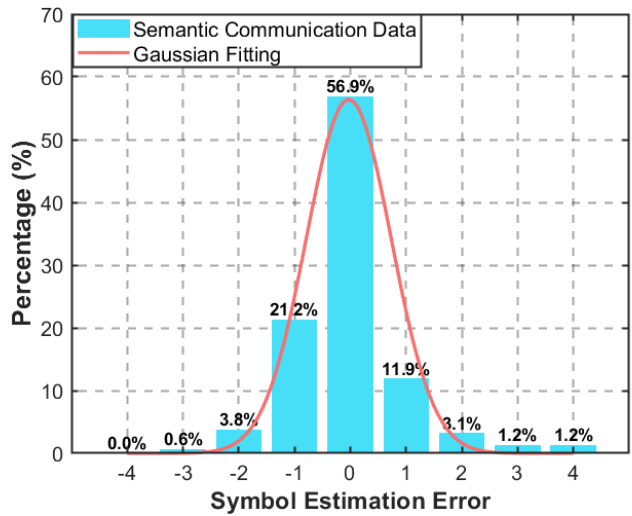

**Fig. 8.** Histogram of symbol estimation errors at 43% symbol error rate. The x-axis represents the deviation between the estimated symbol index and the actual symbol index. And the y-axis represents the proportion of the deviation value. Most estimation errors deviate by only one symbol.

### 4.2.3 Transmission Results under Different Symbol Error Rates

To investigate the effectiveness of our semantic encoding method combined with MMF in achieving accurate semantic transmission under high bit error conditions, we conducted a series of experiments. We transmitted 100 test data samples, which were independent of the training set and unseen by the LSTM network, over a 1 km MMF. By varying the sampling rates, we obtained semantic recognition results under various transmission error rates, as shown in Fig. 9.

As the sampling rate decreases, both the transmission symbol error rate and the sentiment classification error rate gradually increase. However, the overall sentiment classification error rate remains significantly lower than the transmission symbol error rate. This is because when channel degradation causes word errors, our system tends to map the erroneous words to semantically similar words, thus maintaining the overall semantic quality of the transmission and greatly enhancing the system's semantic robustness.

Combining the semantic communication classification network, our system effectively compensates for bit errors during transmission, achieving stronger noise resistance. By adopting the word embedding-based encoding method, we not only enhanced the system's robustness but also enabled efficient transmission within a smaller bandwidth. This approach holds significant potential for practical applications, especially in bandwidth-constrained or noisy communication environments.

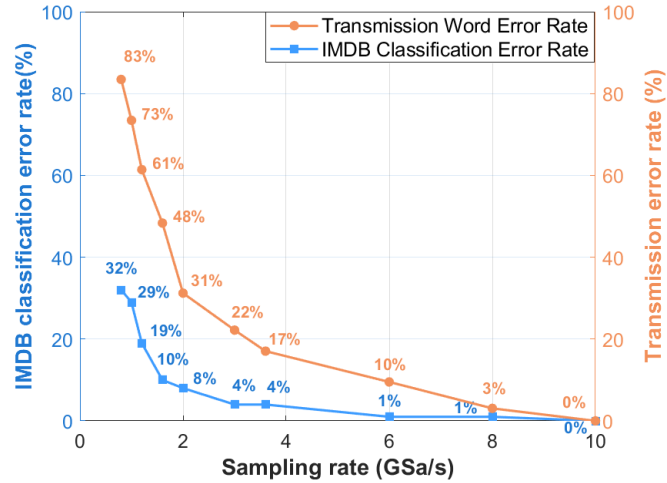

**Fig. 9.** Transmission symbol error rate and IMDb classification error rate at different sampling rates of 1GHz to 10GHz. The orange line represents the transmission symbol error rate at different sampling rates, while the blue line represents the corresponding IMDb sentiment classification error rate. The overall sentiment classification error rate remains significantly lower than the transmission symbol error rate.

*4.3 MCF sampling*

By splicing the seven-core MCF to the MMF, we can sample different positions of the speckle pattern of the MMF, thereby enabling spatial multiplexing of the output signal. Using the previously described encoding and decoding scheme, the MCF allows for the decoding of the same information at seven different receivers. Since the speckle patterns are spatially disordered, the seven intermodal dispersion curves corresponding to the same symbol are different. However, with their respective independent dictionary information, each receiver can achieve correct decoding. The dispersion curves of the MCF are shown in Figure 7. This method enables simultaneous sampling and decoding of the same signal at multiple spatial locations, enhancing both robustness and transmission efficiency.

Experimentally, we demonstrated that although the sampling positions and the signals collected by different cores are entirely uncorrelated, similar reconstruction accuracy can be achieved. Figure 8 presents the symbol error rate for the seven cores at various sampling rates. The reconstruction accuracy and the trend of changes are consistent across all seven cores, achieving zero-error transmission at sampling rates above 9 GHz. When higher reconstruction accuracy is needed to combat noise or further reduce the required spectral bandwidth, the data from several cores can be combined for processing. By analyzing data from all seven cores simultaneously, errors caused by noise were eliminated, achieving zero-error transmission at sampling rates above 6 GHz. This redundancy significantly improves the system's recognition accuracy.

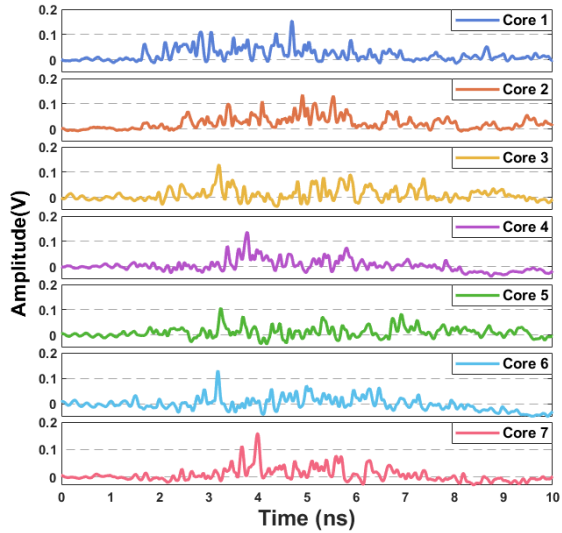

**Fig. 7.** Dispersion curves sampled by the MCF at seven different positions of the MMF speckle pattern. The frequency of the input optical pulses is fixed. The seven curves are uncorrelated since the speckle patterns are spatially disordered.

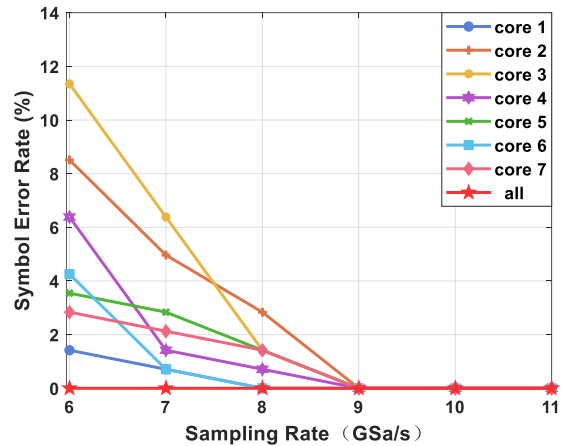

**Fig. 8.** Symbol error rate of the seven cores at sampling rate of 6GHz to 11GHz. Errors are eliminated when analyzing data from all seven cores simultaneously. When the sampling rate exceeds 10 GHz, the symbol error rate for all sampling cores is zero.

## 5. Conclusion

We demonstrated a dense frequency-encoded optical fiber communication system based on MMF. By exploiting intermodal dispersion in MMFs, we create unique frequency fingerprints that allow precise frequency mapping and encoding. We achieved a spectral efficiency of 18.23 bits/s/Hz and a wavelength resolution of 5 fm. The capabilities of intensity modulation and sentiment analysis were experimentally validated. Additionally, the use of a seven-core MCF for spatial multiplexing significantly improved transmission efficiency and robustness. Since the transmission process of the encoded incident signal before entering the MMF does not affect the decoding, this system can be integrated with existing single-mode fiber-based optical communication systems. The signal can first be transmitted over long distances with low loss in single-mode fiber, and then enter the MMF for dispersion-based decoding. This approach holds great promise for future optical communication systems.

**Acknowledgements**

This research is supported by the National Key Research and Development Program of China (2021YFB2800902); National Natural Science Foundation of China (61931010, 62225110); Hubei Province Key Research and Development Program (2020BAA006).

**Conflict of interest**

The authors declare no conflicts of interest.